# Towards the Enhancement of Body Standing Balance Recovery by Means of a Wireless Audio-Biofeedback System

Giovanni Costantini[a], Daniele Casali[a], Fabio Paolizzo[a,b], Marco Alessandrini[c], Alessandro Micarelli[c], Andrea Viziano[c], Giovanni Saggio[a,*]

[a] *Department of Electronic Engineering, University of Rome "Tor Vergata", Via del Politecnico 1, 00133 Rome, Italy*
[b] *Department of Cognitive Sciences, University of California, Irvine 3151 Social Sciences Plaza, Irvine, CA 92697, USA*
[c] *Department of Clinical Sciences and Translational Medicine, University of Rome "Tor Vergata", Via Montpellier 1, 00133 Rome, Italy*

## abstract

Human maintain their body balance by sensorimotor controls mainly based on information gathered from vision, proprioception and vestibular systems. When there is a lack of information, caused by pathologies, diseases or aging, the subject may fall. In this context, we developed a system to augment information gathering, providing the subject with warning audio-feedback signals related to his/her equilibrium. The system comprises an inertial measurement unit (IMU), a data processing unit, a headphone audio device and a software application. The IMU is a low-weight, small-size wireless instrument that, body-back located between the L2 and L5 lumbar vertebrae, measures the subject's trunk kinematics. The application drives the data processing unit to feeding the headphone with electric signals related to the kinematic measures. Consequently, the user is audio-alerted, via headphone, of his/her own equilibrium, hearing a pleasant sound when in a stable equilibrium, or an increasing bothering sound when in an increasing unstable condition. Tests were conducted on a group of six older subjects (59y-61y, SD =2.09y) and a group of four young subjects (21y-26y, SD =2.88y) to underline difference in effectiveness of the system, if any, related to the age of the users. For each subject, standing balance tests were performed in normal or altered conditions, such as, open or closed eyes, and on a solid or foam surface The system was evaluated in terms of usability, reliability, and effectiveness in improving the subject's balance in all conditions. As a result, the system successfully helped the subjects in reducing the body swaying within 10.65%-65.90%, differences depending on subjects' age and test conditions.

## 1. Introduction

The visual, the proprioception and the vestibular systems work to provide standing balance [1,2], so that when one or more of these systems are under sufferance (because of pathologies, diseases, or other reasons) balance capabilities can be reduce, even severely, in effectiveness.

The amount of reduction differs because of number/amount of the systems involved, and can be age-related (because of senescence).

Pathologies of the ear, brain, or sensory nerve can cause dizziness, which affects 47% of men and 61% of women over 70 years of age [3], causing the 25% of falls [4]. Fall-related injuries in older people are a major global health problem, and consequences of falling can psychological produce fear of falling again and depression, which can lead to social isolation [5]. A sedentary lifestyle and drug usage can induce a slowdown of reflexes even in young people [4] (statistics not yet available).

When the sense of balance is reduced because of system issue(s), subjects tend to compensate by means of other systems, as it can be when information from the vestibular system increases in importance if the visual system is limited or absent. Key enabling technology (KET) can provide the subject with information no more or insufficiently supplied by the vestibular system [6–10]. Within this frame, here we propose an audio-biofeedback-based technology (ABF-T) as a KET useful to provide subjects with audio signals coded on the basis of his/her measured standing balance.

The ABF-T consists of wearable devices, or wearables, which are a belt-worn inertial measurement unit (IMU) plus a headphone, and desktop hardware, which comprises a data receiving station and a personal computer. In particular, the wearables are battery-powered low-weight, low-size, operator-independent, and low-cost apparatuses. The desktop hardware, wireless communicating with the wearables, provides A/D and D/A conversions, and real-time process data feeding signals to the headphone.

* Corresponding author. *E-mail address:* saggio@uniroma2.it, wmdtg@email.it (G. Saggio).

The overall system is validated with ten subjects, grouped in young and older people, performing four different condition tests, so to take into account of normal, and altered-visual (i.e. with closed eyes), and of altered-proprioceptive (i.e. on a foam surface) conditions.

## 2. State of the art

Within biofeedback systems [11–14], ABF is progressively acquiring greater relevance [15,16]. For example, ABF has been successfully exploited for motor rehabilitation after a hemiparetic stroke [17], when a patient was asked to move the hand to virtually displace an object represented on a computer screen. A cone-like surface changed in color and a sound turned into musical notes according to the right or wrong trajectory of the patient's hand. As a result, patients with sensory-motor deficits better performed compensatory movements using that system. In a recent study, Parkinson's patients enhanced their balance by using an ABF-based therapy [18]. According to Petrofsky [19], ABF was used for the therapy of five subjects with impaired walking capability and reduced muscle strength of the hip adductor (Trendelenburg's sign), caused by spinal cord injury. By means of EMG sensors, the activity of the gluteus medium muscle was measured, and the patients were provided with audio signals coded to inform of incorrect posture. The ABF was implemented successfully also by Dozza et al. [20], who performed tests on healthy subjects in standing position, with closed eyes and on foam on top of a force plate. The subjects' balance information was related to stabilogram diffusion and center of pressure analysis, from which a signal was conceived and audio-provided in a way that the subjects' postural stability increased.

Our ABS system evaluates the swaying characteristics of standing subjects on a solid surface and on a foam surface, in both conditions of open and closed eyes. Innovatively, we empirically defined different regions of swaying and audio-coding algorithms related to those regions, other from previously reported [21,22].

## 3. Materials and methods

The ABF system is based upon three logical components: a *sensor unit* (consisting of an inertial measurement unit, IMU, device) that measures the subject's movements, a *processing unit* (consisting of an electronic device and a personal computer) that codes the sensory information in real time (routines written in Max/MSP, a Max Software Tools for Media, by Cycling '74, San Francisco, USA), and an *output unit* (consisting of a headphone) that provides the subject an acoustic representation of the physiological parameters.

### 3.1. Hardware

The sensor unit is a belt-worn low-size (4.77 × 4.12 × 1.76 cm), low-weight (20 g), scientifically validated [23] IMU, termed Movit (Fig. 1(a), by Captiks Srl, Rome, Italy), which provides inertial data related to the movements of the subject. The Movit is operated by a 40 MHz clock microcontroller (AT32UC3A4256 by Atmel, San Jose, California, US) and includes a 3D gyroscope, a 3D accelerometer, a 3D compass (each sensors integrated in a MEMS MPU-9150 by Invensense, San Jose, California, US), and a barometer (BMP-180 by Bosch Sensortec, Reutlingen, Germany). The Atmel microcontroller wireless sends data via 802.15.4 protocol, and a built-in Li-Po battery provides up to 8-hour supply, thus allowing no interruptions during tests (of only few minutes in our case).

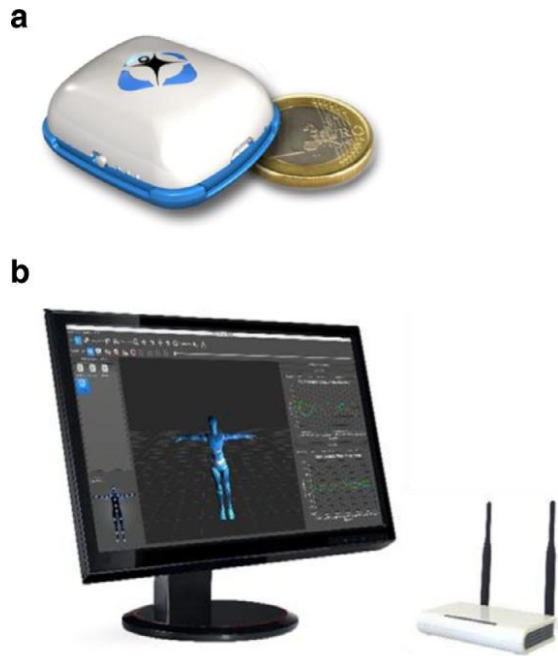

**Fig. 1.** (a) The Movit unit, dressed by the subject, which integrates a 3D gyroscope, a 3D accelerometer, a 3D compass, a barometer, and a Li–Po battery, (b) the receiving unit, and the software termed "Captiks Motion Studio SDK, CMS".

The Movit unit (which can be networked with up to 15 other units if necessary), wireless sends data to a receiving device (the *receiver*) connected to a personal computer. Specifically, we acquired data from the internal 3-axis digital-output accelerometer, which has a programmable full-scale range of ±2 g, ±4 g, ±8 g and ±16 g, and integrates a 16-bit A/D converter, enabling simultaneous sampling of accelerometers without requiring an external multiplexer. Data rate can range within 4 Hz–1000 Hz. We operated the unit in low-power mode, at an operating current of 140 $\mu$A with a 50 Hz updating rate. The Movit was housed in a belt and located at the subject's spine level, between the second and the fifth lumbar vertebra (L2 and L5, Fig. 2). We used a subset of two (out of three) parameters of the accelerometer, as useful for our purposes, in particular the pitch angle ($x$ coordinate, known as AP: Anterior/Posterior), and the roll angle ($y$ coordinate, known as ML: Medial/Lateral).

A personal computer (i5 processor by Intel, 4GB Ram), connected to the receiver component, converts data into coded sounds furnished to the subject by means of headphone (model K701 by AKG) with large-band frequency response (10Hz-39.800 Hz) and high dynamics (105 dB SPL/V). The headphone is an "open-type", which allows the subject hearing the voice of an assistant, if necessary.

### 3.2. Software

The personal computer runs proprietary software (termed Captiks Motion Studio SDK, CMS, by Captiks Srl) for motion analysis (Fig. 1(b)), so that pitch ($x$) and roll ($y$) signals (Fig. 3) are A/D converted and further normalized, within 0–1 range. Output data are converted into proper sound signals by means of routines based on Max/MSP (which is a graphical dataflow programming environments for audio, by Cycling '74, San Francisco, USA) [24].

### 3.3. Regions and ranges of swaying

We modelled the body of the subject as an inverted pendulum, so that pitch ($x$) and roll ($y$) of the trunk motion shaped a 2D orthogonally projected *point P(x,y)* on the floor (the coordinates of *P* expressed in degrees). By means of Max/MSP routines we generated two-channel signals related to the coordinates of the point *P(x,y)*, accordingly to audio coding as defined in the following (Section 3.4).

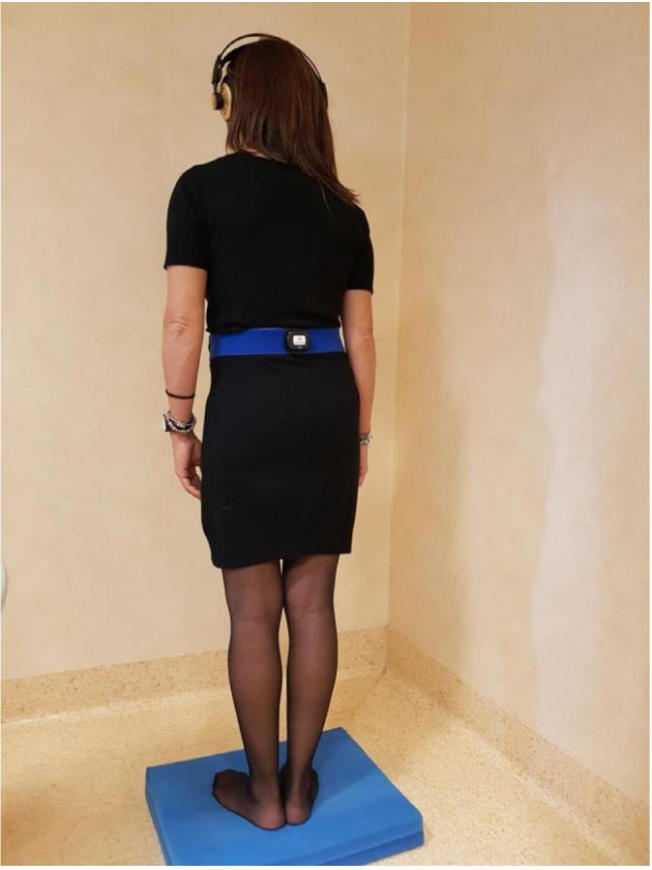

Fig. 2. The Movit was located at the subject's spine level, between L2 and L5. Here, the subject stands on a rubber foam with closed eyes.

We consider the *swaying* as the swing back-and-forth or to- and-from, i.e. the pitch and roll angles, as a distance *dist* of *P(x,y)* with respect to the position of *P(x_o,y_o)* itself when the subject naturally stands during the calibration phase, $dist = P(x,y) - P(x_o,y_o)$, where $x = x(t)$, $y = y(t)$, and so $dist = dist(x,y,t)$.

Accordingly, *P(x,y,t)* can result within six different regions (from "A" to "F") that we empirically identified in the 2D floor-plane, as schematized in Fig. 4.

For each test (later discussed in Section 3.6) and for each partecipant (described Section 3.5), we determined the *range R*, as the difference between maximum and minimum distance *dist(x,y,t)*, and the related *variance V*, both in absence and in presence of ABF ($R_{noABF}$, $R_{ABF}$ and $V_{noABF}$, $V_{ABF}$, respectively). Finally, $P_R$ and $P_V$ express such variations as percentages, as follows:

$$P_R = [(R_{noABF} - R_{ABF})/R_{noABF}] \cdot 100 \quad (1)$$

$$P_V = [(V_{noABF} - V_{ABF})/V_{noABF}] \cdot 100 \quad (2)$$

Accordingly, $P_R$ represents normalized differences between values of the *Range* obtained without and with the ABF-T, respectively, so that the higher $P_R$ the more effectiveness of the system.

The regions of swaying identify "safety" (region "A"), "low-level warning" (region "B"), "medium-level warning" (region "C"), and "high-level warning" (regions "D", "E", "F") conditions, and the audio-sounds are generated accordingly, in order to "reassuring" or "softly/hardly alerting" the subject. These regions are delimited by contours described by mathematical functions of circle, ellipse, and rectangle, as reported in Table 1.

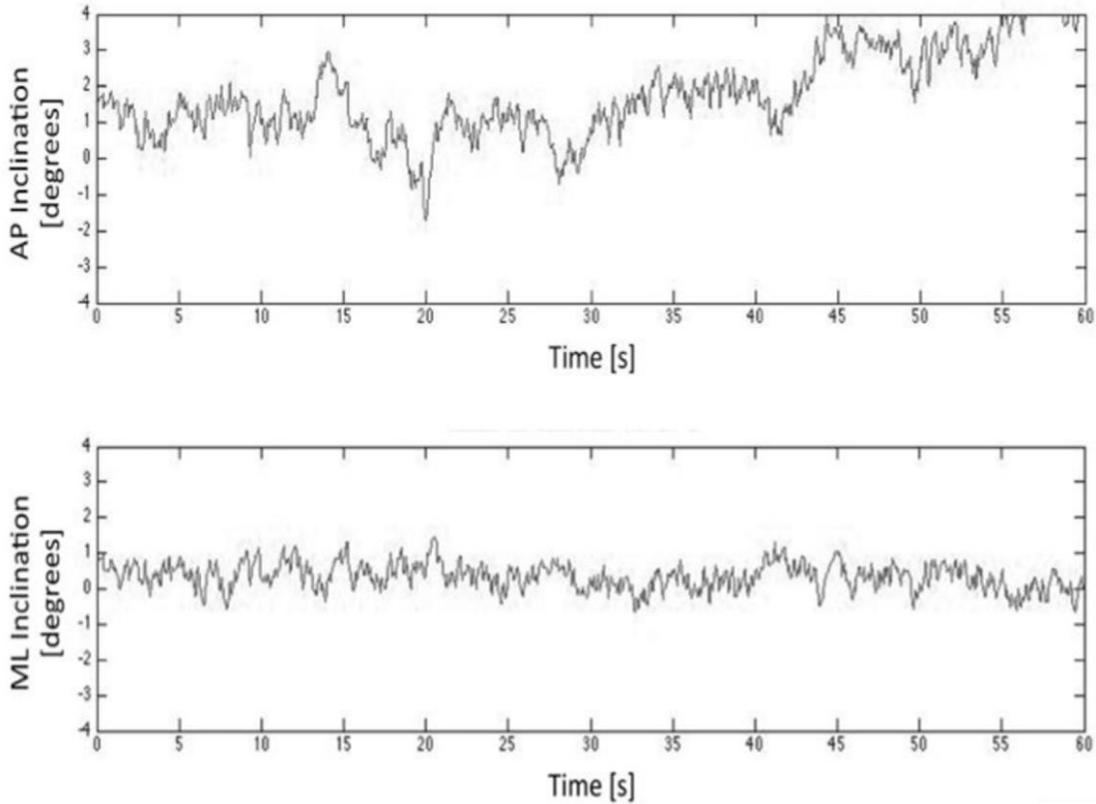

Fig. 3. Pitch vs. time (a) and roll vs. time (b) for subject 2, with closed eyes, on foam rubber and without ABF.

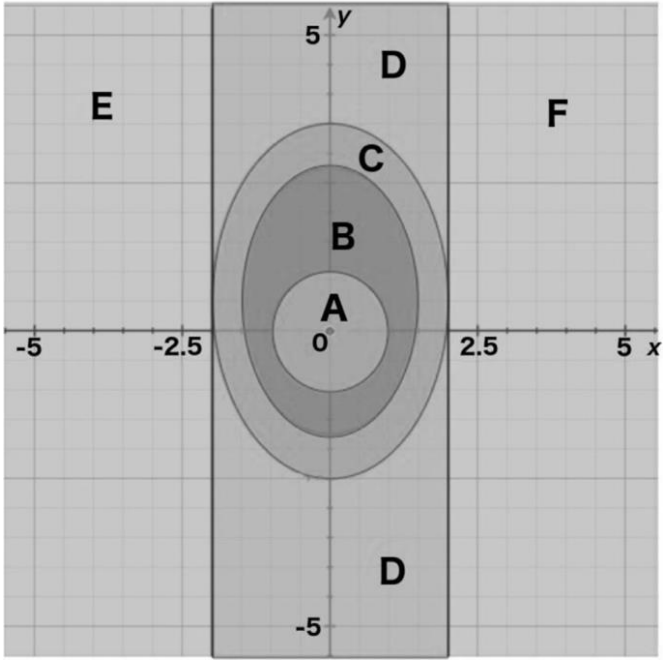

**Fig. 4.** Schematization of the six 2D different regions where the pitch (*x*) and roll (*y*) signals, coming from the IMU, are projected as a point *P*. Regions are divided among A: safety, B: low-level warning, C: medium-level warning, D, E, F: high-level warning, different in signal-balance between hears.

**Table 1**
Identification of the six regions of swaying, from "A" to "F", by means of their mathematical functions and levels of warning.

| Region | Contour | Warning? |
|---|---|---|
| "A" | $x^2 + y^2 = 1$ | No (safety) |
| "B" | $[(y-0.5)/2.25]^2 + (x/1.5)^2 = 1$ | Low-level |
| "C" | $[(y-0.5)/3]^2 + (x/2)^2 = 1$ | Medium-level |
| "D" | $-2 < x < 2$ and outside Region 3 | High-level |
| "E" | $x \leq -2$ | High-level |
| "F" | $x \geq 2$ | High-level |

Regions "B" and "C" are neither circular nor centered, since frontal tolerance of swaying is greater than backward or lateral tolerance of swaying in body balance (because of foot shape).

### 3.4. Audio coding

Audio is coded from pink noise to an almost pure sine wave (Fig. 5 shows the intermediate occurrences), according to the severity of the unbalance status, and the sounds between the two hear channels are stereophonically balanced or unbalanced in amplitude.

In particular, according to the pitch values, when $x < 0$ ($x > 0$) the amplitude of the left channel increases at a given gain $g(x)$, while the amplitude of the right channel decreases, and vice versa, by the same amount. Audio coding was different for each region, as detailed in the following:

- When the point $P(x,y,t)$ is within Region "A", the subject is considered to have a correct posture, with no risk of losing his/her balance. In such a condition, we considered that headphone's silence could induce the perception of an unrealistic and annoying "absence of assistance", which could even produce ambiguity between "no warning" and "no working of the system." On the other end, a pure sine tone was bothersome for many participants. Therefore, similarly to telephone company providers that generate a "courtesy noise", we implemented a pink noise generator, which informs the subject that everything is correctly working without alerts. The reference volume of the pink noise is set by the user him/her-self, according to preference.
- When the point $P(x,y,t)$ is within Region "B", a "low-level warning" alerts the subject by an audio noise empirically filtered within 128Hz-14,263 Hz range, with a volume 1.5 higher than the reference.
- When $P(x,y,t)$ is within Region "C", an alerting noise is generated, band-pass filtered within 415Hz-4390 Hz, with a volume three times higher than the reference.
- When $P(x,y,t)$ is within Region "D", an 800 Hz narrow-band filtered noise is generated. The frequency band is within a lower cut-off frequency $f_{inf}$ empirically set as an exponential function of *y*, $f_{inf} = 2^{((8+(y+20)/40) \cdot 4)}$, and a higher cut-off frequency sets as $f_{sup} = f_{inf} + 800$ Hz. In such a way, we provide the subjects with a pitch sound variation proportional to the tilt angle. With such a narrow band-pass filter, the subject perceives the sound as an almost pure sinusoidal wave. Based on the logarithmic human perception of pitch, we designed this perception to become stronger with frequency increasing.
- When $P(x,y,t)$ is within Region "E" and "F", we used the same approach as for Region "D", i.e. 800 Hz narrow-band filtered noise, but modulated by a square wave panned to the headphone channels (left, $x \leq 2$, for Region "E" and right, $x \geq 2$, for Region "F"), in order to obtain an intermittent noise. The square wave ranged from 0 to 1 at a duty cycle of 50%, with the period *T* exponentially decreasing in function of $|x|$, namely $T = 8^h$, with exponent $h_{|x|=0} = 2.5$, $h_{|x|=20} = 2$, so that $T_{min} = 82$ ms, $T_{max} = 82.5$ ms (*h* goes linearly from 2.5 to 2 when $|x|$ changes from 0 to 20). The resulting train of pulses had a frequency of about 15.6 Hz when $|x| = 20$, which also corresponds to the maximum considered angle.

Fig. 6 shows region-related cut-off frequencies. All signals were wireless provided to the subjects so avoiding cables which could obstruct the body swaying.

### 3.5. Participants

In order to compare balancing abilities between groups [25], tests were conducted on 10 healthy subjects, divided into groups of young and older people. Specifically, four young subjects (21y-26y, SD = 2.88y) and six elder subjects (59y-61y, SD = 2.09y), equally divided into male and female. The study was approved by the local ethics committee, and all participants gave their informed consent in writing.

### 3.6. Test protocol

According to a previously validated protocol [8], the balance of each subject was considered in four different test conditions while standing upright: (1) on floor with opened eyes; (2) on floor with closed eyes; (3) on rubber foam layer with opened eyes; (4) on rubber foam layer with closed eyes. Rubber foam layer and closed eyes conditions were considered to test deprived sensory conditions so simulating impairment at the proprioception level [26]. In particular, the rubber foam is a 6-cm heightened AIREX Balance-pad plus foam carpet (FC; Airex AG, Sins, Switzerland) 25% compression resistance, 20 kPa; apparent density 55 kg m3; tensile strength 260 kPa).

All tests were performed in a quiet room, at a room temperature of 25 °C, with the subject, in turn, quietly with head still, while maintaining the upright standing position for 60 s. Each subject was requested to reduce his/her body natural oscillation to a minimum. The four conditions were tested both with and without ABF, so resulting $4 \times 2 = 8$ (#tests x #replies) sessions for each subject, while the sampling acquisition data rate was 50 Hz, so resulting a total of (50 Hz x 60s = ) 3000 samples for each session.

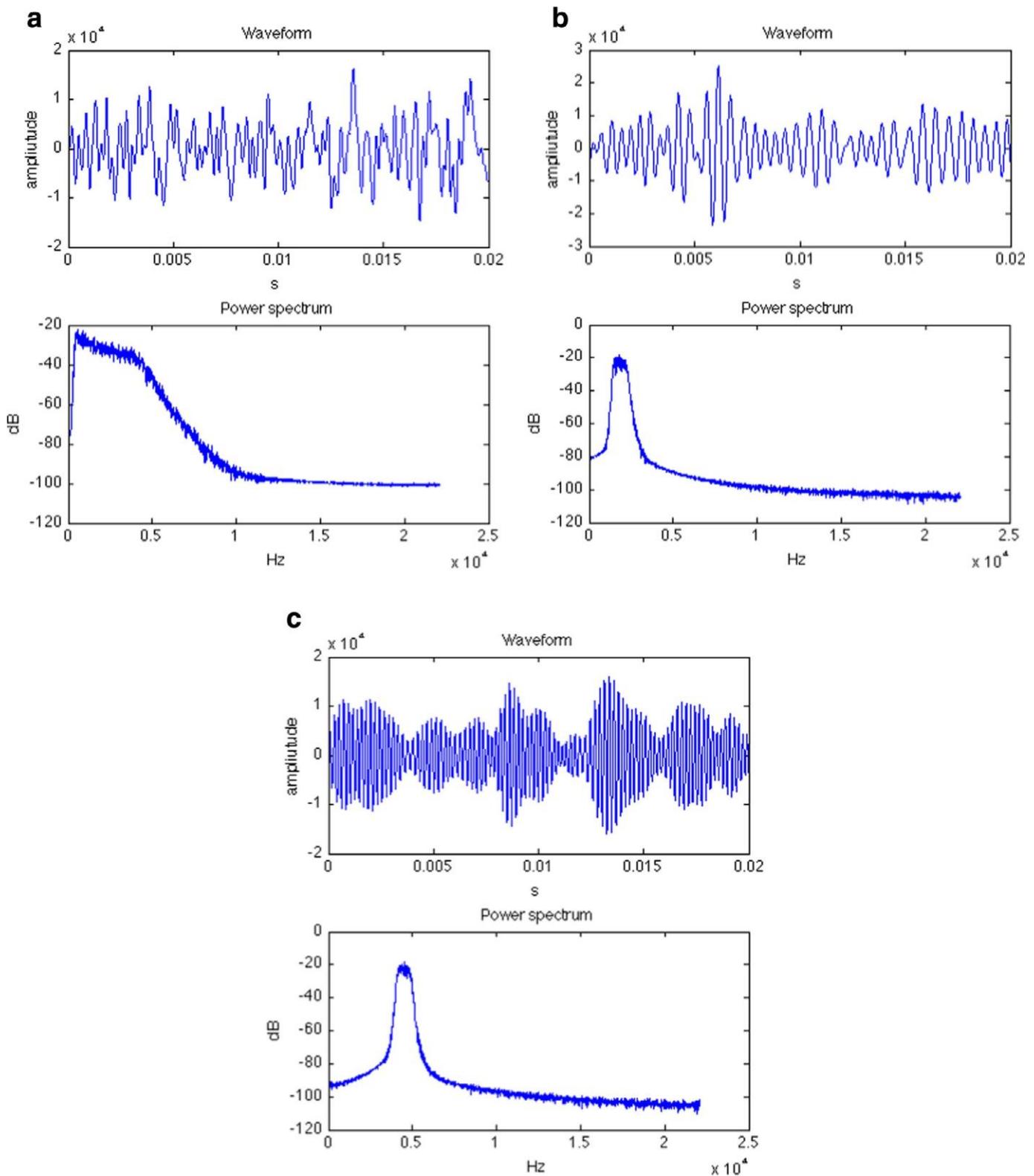

**Fig. 5.** (a) Waveform and amplitude spectrum of the sound generated when (a) $x = 0$; $y = 3$ (filtered pink noise), (b) $x = 0$; $y = 5$ (mid freq. almost pure sine), (c) $x = 0$; $y = 20$ (high freq. almost pure sine).

Before performing, each subject was trained for some minutes to get confident with the protocol.

## 4. Results and discussion

Fig. 7 eloquently shows the way point ***P*** "disperses" in eight different occurrences of tests related to one older subject, combining

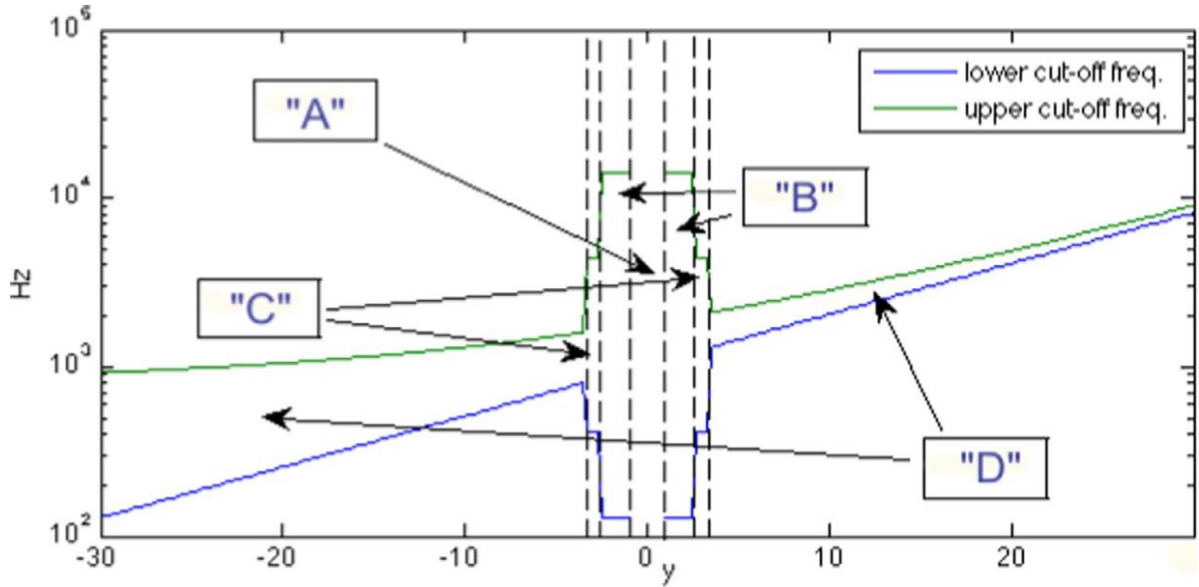

**Fig. 6.** Upper and lower cut-off frequencies vs. pitch, according to the regions.

**Table 2**
$P_R$ and $P_V$ values resulted from four different test conditions while subjects were, in turn, standing upright. Conditions were: on floor with opened eyes; on floor with closed eyes; on rubber foam layer with opened eyes; on rubber foam layer with closed eyes.n.

| Eyes open or closed | floor or foam | Older group $P_R$ [%] | $P_V$ [%] | Younger group $P_R$ [%] | $P_V$ [%] | Overall $P_R$ [%] | $P_V$ [%] |
|---|---|---|---|---|---|---|---|
| 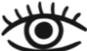 | 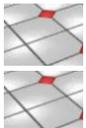 | 10.65 | 29.63 | 56.04 | 65.55 | 26.92 | 49.24 |
| 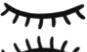 | 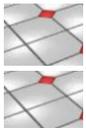 | 34.44 | 58.10 | 53.53 | 64.74 | 40.62 | 58.10 |
| 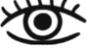 | 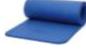 | 31.97 | 45.76 | 49.21 | 65.90 | 35.16 | 50.68 |
| 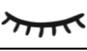 | 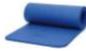 | 29.87 | 26.19 | 38.27 | 46.23 | 36.61 | 37.86 |

situations of open/closed eyes, floor/foam platform, with/without ABF.

The ABF always increased the body standing balance, so that the differences of ($R_{noABF}$ $R_{ABF}$) and ($V_{noABF} - V_{ABF}$) always result in a positive value. $P_R$ (Eq. (1)) and $P_V$ (Eq. (2)) result positive, consequently. The increase in body balance, measured as the median values of $P_R$ and $P_V$, were computed for the elders (by means of home-made routines written in Matlab R2017b, by Mathworks, Massachusetts, US), for the youngers and for the ensamble of the two groups, as schematized in Table 2.

According to our results, the increase in body balance is greater for young than older subjects, but meaningful in both cases. Specifically, the presence of the ABF improved the performance under conditions of closed eyes and standing on floor for the older group ($P_R$ = 34.44, $P_V$ = 58.10), and conditions of opened eyes and standing on floor for the young group ($P_R$ = 56.04, $P_V$ = 65.55). Differently, the ABF in conditions of opened eyes and standing on floor was less effective for the young group ($P_R$ = 10.65, $P_V$ = 29.63). Similar results are reported, among the same group, for all of the other conditions.

When standing on the floor, the loss in visual feedback (closed-eyes condition) makes the ABF useful in any case, but more relevant for older subjects than for young people (from $P_R$[%] = 10.62 to $P_R$[%] =34.44, and from $P_R$[%]=56.04 to $P_R$[%]=53.53, respectively).

When standing on the foam, the ABF results more effective for the young group when combined to the visual stimulus, the latter being inborn with respect the few-minutes trained ABF ($P_R$[%] = 49.21 for open-eyes and $P_R$[%]=38.27 for closed-eyes).

**5. Conclusions**

We designed an ABF-T based system and tested it, both on a group of six older people and on a group of four young subjects. The system was able to measure the body swaying and to generate a sound corresponding to the measurements. This was to alert the subject that his/her own balance was unstable with the potential consequence of falling.

The results demonstrated the effectiveness of ABF in helping the subject in maintaining body balance, under different test conditions. The distance of swaying in angle from the origin, namely the *range R*, and its *variance V*, were always lower when the ABF was used with respect to their counterpart than when the ABF was absent. These occurrences were confirmed in all the different test conditions: four combinations of open/closed eyes and with/without foam rubber under the subject's feet.

As a result, the ABF is capable to provide the subject with additional information about his/her own somatosensory system to better control the body movements. We reported the worst results in closed-eyes condition with foam rubber under the feet, as the total absence of visual and somatosensory information reduced the subject's capacity to control his/her own movements. Both the open-eyes condition with foam rubber and the closed-eyes condition without foam rubber produced intermediate results. In conditions of closed eyes with the use of foam rubber, the system proved to be more effective for the young subjects with respect to the older group of people.

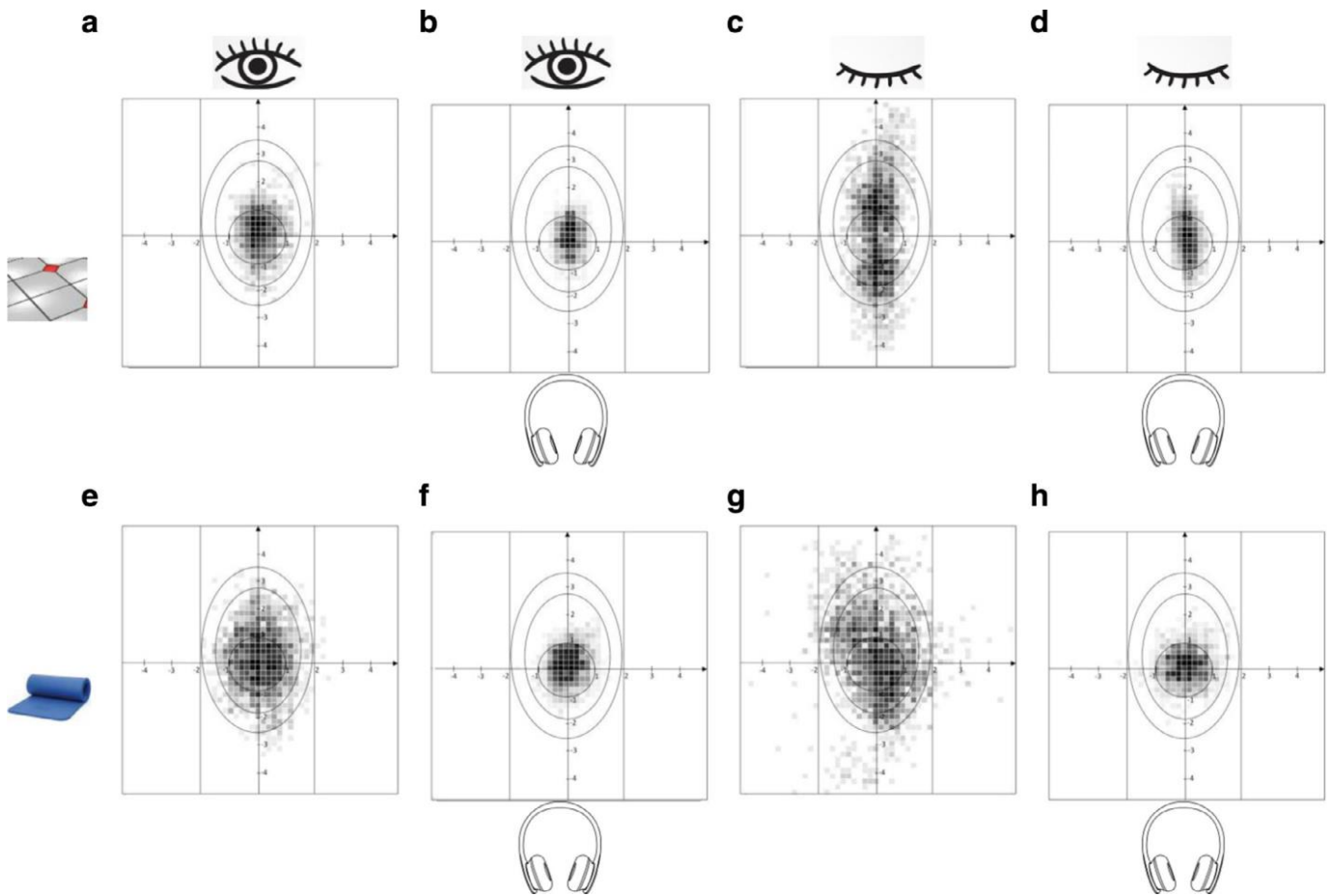

**Fig. 7.** An exemplification of the "dispersion" of the point *P* on the base-plane related to the swaying assumed by one elderly subject when upright standing: (a) on floor, opened eyes, no ABF; (b) on floor, opened eyes, with ABF; (c) on floor, closed eyes, no ABF; (d) on floor, closed eyes, with ABF; (e) on foam, opened eyes, no ABF; (f) on foam, opened eyes, with ABF; (g) on foam, closed eyes, no ABF; (h) on foam, closed eyes, with ABF.

The ABF positively and differently augmented the balance recovery in every test condition and for every subject. Anyway, we argue that the higher effectiveness of the system on young people can be due to their higher reactiveness to external feedbacks, and quicker response to multimedia stimuli (as, for instance, videogames can be). It can be also possible that a training period of few-minutes only, adopted to get confident with the system, played a rule too, since older people may claim a longer time to get used to ABF.

Finally, the ABF technology represents a low-weight, low-cost and non-intrusive system capable to enhance the subject's standing balance. These characteristics suggest that ABT-T can augment traditional rehabilitation practice in a clinical setting or increase compliance in at-home-based postural exercise programs, successfully.

**Acknowledgements**

*Conflict of interest*
We declare no competing interests.

*Ethics declarations*
The study was approved by the local ethics committee (Comi- tato Etico Indipendente, Tor Vergata, Protocollo 35/17).

*Funding*
The research was partially funded by the European Union's Horizon 2020 research and innovation programme under the Marie Sklodowska-Curie grant agreement No 659434.

**Supplementary materials**

Supplementary material associated with this article can be found, in the online version, at doi:10.1016/j.medengphy.2018.01.008.